\documentclass[superscriptaddress,longbibliography,twocolumn]{revtex4-2}
\usepackage[utf8]{inputenc}
\usepackage{color}
\usepackage{graphicx}
\usepackage{array}
\usepackage{complexity}
\usepackage[dvipsnames]{xcolor}
\usepackage{tcolorbox}
\usepackage{tikz}
\usetikzlibrary{positioning}
\usepackage{hyperref}
\usepackage{appendix}
\hypersetup{
    colorlinks=true,
    linkcolor=blue,
    filecolor=magenta,      
    urlcolor=cyan,
    citecolor=blue, 
    pdfpagemode=FullScreen,
    }
\usepackage{ragged2e}
\usepackage{multirow}

\definecolor{our-red}{RGB}{248,173,150}
\definecolor{our-blue}{RGB}{170,205,233}
\definecolor{our-lightblue}{RGB}{214,241,247}
\definecolor{our-yellow}{RGB}{234,225,95}
\definecolor{our-lightyellow}{RGB}{255,241,170}
\definecolor{our-lightgray}{RGB}{226,226,226}
\definecolor{our-gray}{RGB}{206,212,219}

\begin{document}

\title{Drug design on quantum computers}
\newcommand{\Google}{\affiliation{%
Google Quantum AI, Venice, CA 90291, United States}}

\newcommand{\Toronto}{\affiliation{%
Department of Computer Science, University of Toronto, Canada}}

\newcommand{\BI}{\affiliation{%
Quantum Lab, Boehringer Ingelheim, 55218 Ingelheim am Rhein, Germany}}

\newcommand{\BIMedChem}{\affiliation{%
Boehringer Ingelheim Pharma GmbH \& Co KG, Birkendorfer Strasse 65, 88397 Biberach, Germany}}

\newcommand{\UIBK}{\affiliation{Department of General, Inorganic and Theoretical Chemistry,
University of Innsbruck, 6020 Innsbruck, Austria}}

\newcommand{\UVIE}{\affiliation{Institute of Theoretical Chemistry, Faculty of Chemistry, University of Vienna, Währinger Straße 17, 1090 Vienna, Austria}}

\newcommand{\QCW}{\affiliation{QC Ware Corp, Palo Alto, CA 94306, United States}}

\newcommand{\BASF}{\affiliation{Next Generation Computing in Global Digitalization, BASF SE, Carl-Bosch-Strasse 38, 
67056 Ludwigshafen am Rhein, Germany}}

\author{Raffaele Santagati}
\thanks{raffaele.santagati@boehringer-ingelheim.com}
\BI

\author{Alan Aspuru-Guzik}
\Toronto

\author{Ryan Babbush}
\Google

\author{Matthias Degroote}
\BI

\author{Leticia Gonz\'{a}lez}
\UVIE

\author{Elica Kyoseva}
\thanks{Present Address: Wellcome Leap Inc., Los Angeles, CA 90069, United States}
\BI

\author{Nikolaj Moll}
\BI

\author{Markus Oppel}
\UVIE

\author{Robert M.~Parrish}
\QCW

\author{Nicholas C.~Rubin}
\Google

\author{Michael Streif}
\BI

\author{Christofer S.~Tautermann}
\BIMedChem
\UIBK

\author{Horst Weiss}
\BASF

\author{Nathan Wiebe}
\Toronto

\author{Clemens Utschig-Utschig}
\BI

\begin{abstract}
Quantum computers promise to impact industrial applications,  for which quantum chemical calculations are required, by virtue of their high accuracy.
This perspective explores the challenges and opportunities of applying quantum computers to drug design, discusses where they could transform industrial research and elaborates on what is needed to reach this goal.
\end{abstract}
\maketitle

\section*{Introduction}
For over fifty years, the pharmaceutical industry has seen the cost of developing drugs increase exponentially from tens of millions in the 1950s to billions of dollars today, even when the data is adjusted for inflation~\cite{Scannell2012}.
To sustain the progress in treating unmet medical need, it is essential to look for every source of improvement in the methodologies employed in drug development.
In the last decades, computational approaches started to play an increasingly large role in research and development~\cite{Allen2022, Palermo2014}. Many computational methods are employed from machine learning~\cite{maltarollo2015applying, Jayatunga2022} and molecular dynamics~\cite{Shukla2021,Irle2020} to quantum mechanical calculations~\cite{Heifetz2020}. Still, simulating chemical systems, including quantum mechanical effects, can be computationally intensive and many of these methods face limited practical applicability because of speed and accuracy.

By exploiting their quantum mechanical properties, quantum computers have been proposed to simulate quantum systems efficiently~\cite{Feynman1982, Lloyd1996, Nielsen2000, Aspuru2005, Cao2019}. Inspired by this promise, quantum computing research has proliferated in recent years, and a community of quantum physics, chemistry, and information theory experts has brought improvements in quantum hardware and algorithms~\cite{Bauer2020, Liu2022}. The recent developments also attracted interest beyond academia to find practical applications in industry, with investments from private and public sectors. Often, one of the justifications for those investments is the promise that quantum computers will enhance quantum chemistry calculations~\cite{Liu2022, Motta2022, Zinner2021, Baiardi2022, Blunt2022}. Most current efforts in quantum computing focus on finding quantum algorithms for the most challenging electronic structure problems, for which the largest possible advantage over classical computations can be expected. However, identifying such systems with strong electronic correlations is difficult~\cite{Garnet2022}, and there only are a limited number of indicators, such as those shown in Box~\ref{BOX:GoodCandidatesQC}. While solving the electronic structure problem is an important step for many chemical applications, if the advantage of quantum computers is limited to strongly correlated systems, they might have limited practical significance in drug design. 
\begin{table}[htb!]
\refstepcounter{table}
\begin{tcolorbox}[colback=our-lightblue,colframe=our-blue,coltitle=black,title=Box \thetable: Some indicators of strong electronic correlation]
\justifying\noindent Quantum computers are expected to offer an advantage for solving the electronic structure problem of strongly correlated systems. Five different indicators with their graphical representations are shown. There are two regions labelled {\sf Classical} for cases solvable on a classical computer~\cite{bofill1989unrestricted, Khedkar2021, Andersson1990, angeli2001, Pernal2018} and {\sf \color{red} Quantum} for cases where a quantum computer might be required.
\begin{tabular}{p{40mm} p{30mm}}
\vspace*{-10mm}
\textbf{Multi-reference:} system's wavefunction  requiring many reference states (determinants) with comparable amplitudes~\cite{Coe2015,Li2019,Goings2022}. &
\parbox[c]{30mm}{\includegraphics[width=0.5\textwidth]{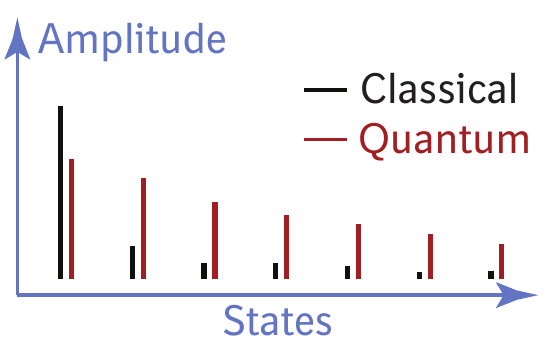}}\\ \vspace*{-10mm}
\textbf{Essential spin-symmetry breaking:}  not fixed by 
adding dynamical correlation~\cite{lee2019distinguishing}. &
\parbox[c]{30mm}{\includegraphics[width=0.5\textwidth]{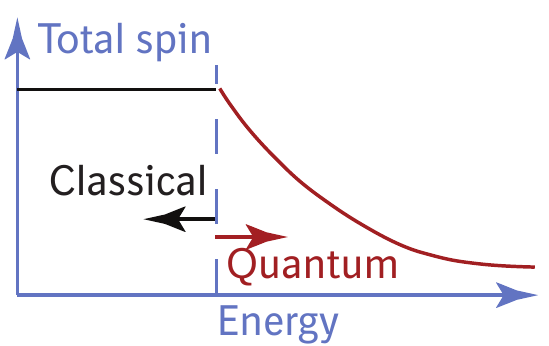}}\\ \vspace*{-10mm}
\textbf{Cluster expansion} have characteristic failure points indicating the need for a multi-reference model~\cite{cheng2017bond,degroote2016polynomial}. &
\parbox[c]{30mm}{\includegraphics[width=0.5\textwidth]{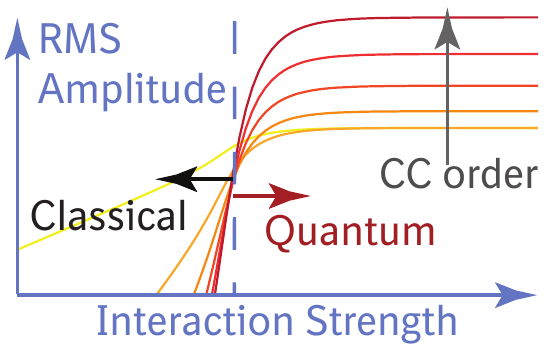}}\\ \vspace*{-10mm}
\textbf{Near degenerate natural orbitals} with non-integer occupation numbers, e.g.\ detected from orbital occupation analysis~\cite{rissler2006measuring,stein2016automated,bofill1989unrestricted}. &
\parbox[c]{30mm}{\includegraphics[width=0.5\textwidth]{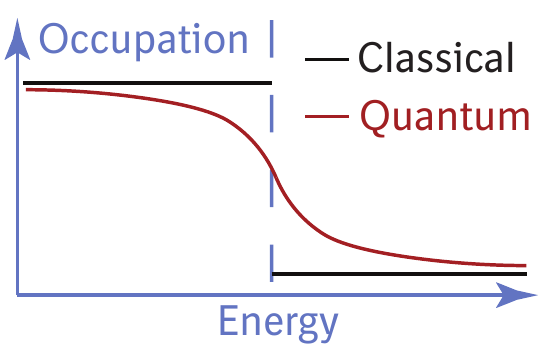}}\\ \vspace*{-10mm}
\textbf{The number of entangled orbitals} grows proportionally to system size, which also needs to be large enough to be classically hard~\cite{chan2011density,Stein2017}, image adapted from~\cite{Ding2021}.&
\parbox[c]{30mm}{\includegraphics[width=0.5\textwidth]{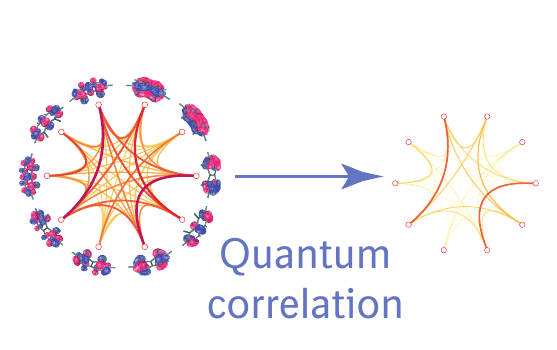}}
\end{tabular}
\label{BOX:GoodCandidatesQC}
\end{tcolorbox} 
\end{table}

In this perspective, we discuss the status quo of the applicability of future quantum computers to problems in drug discovery; specifically, we focus on quantum chemistry calculations because, in our opinion, these will be the first viable applications to impact drug design. 

While we do not give an exhaustive presentation of the status of quantum computing, we discuss problems in quantum chemistry for which quantum computers could offer a speed-up compared to classical computing methods and compare these problems with the actual computational needs in computer-aided drug design. Lastly, we discuss research directions to make quantum computers an essential tool in the pharmaceutical industry.

\section{Status Quo: quantum computers}

The field of quantum computing has seen rapid developments in the last decade~\cite{Liu2022, Cao2019, Bauer2020}. Still, the way towards a practical quantum advantage requires major progress for hardware and algorithms~\cite{Motta2021, McArdle2020}. The most important metric for the development of quantum algorithms is the estimation of their computational cost. These estimates define the quantum computing resources (qubits and run-time) required to solve a problem of interest. They provide concrete engineering targets for quantum hardware and shed light on what aspects of the algorithms need improvements.

Today, only Noisy Intermediate Scale Quantum (NISQ) computing hardware exists, named after its noisy nature and the limited number of qubits~\cite{Preskill2018, Bharti2022}. Most NISQ algorithms, e.g., variational quantum eigensolvers (VQE)~\cite{McClean2014, Cerezo2021}, heavily rely on classical optimisation heuristics, and the actual run-time is difficult to estimate. 
Also, recent results suggest that in NISQ, the number of measurements required to achieve a given error scale exponentially with the depth of the circuit~\cite{Quek2022}. For these reasons, we focus our discussion exclusively on fault-tolerant quantum computers (FTQCs). 

FTQCs exploit quantum error correction to exponentially suppress errors~\cite{Campbell2017}, at the cost of considerable additional qubits and run-time. 
For example, simulating a classically challenging molecule, such as the iron-molybdenum complex (FeMoco)~\cite{Reiher2017}, would require roughly 200 logical (error-corrected) qubits which would be implemented in 2 million physical qubits~\cite{Lee2021}, well beyond what is achievable with current quantum hardware~\cite{Bharti2022}. 

Quantum computers are expected to offer a clear advantage in finding the ground state energy of a molecular Hamiltonian (i.e.\ solving the electronic structure problem) for strongly correlated systems where all tractable classical methods fail.
To identify those systems, several conditions need to be satisfied (see Box~\ref{BOX:GoodCandidatesQC}), and verifying them can be very demanding and time-consuming and heavily relies on chemical expertise. Over the past twenty years, several techniques have been developed for studying how and when various \textit{ab initio} methods fail, delivering indicators of strong correlations~\cite{Stein2017}. Typical examples of such situations that require expensive multi-reference treatment are multi-metal systems, where metals are in similar electronic environments and interactions. 

\begin{figure*}[htb!]
    \centering
    \includegraphics[width=\textwidth]{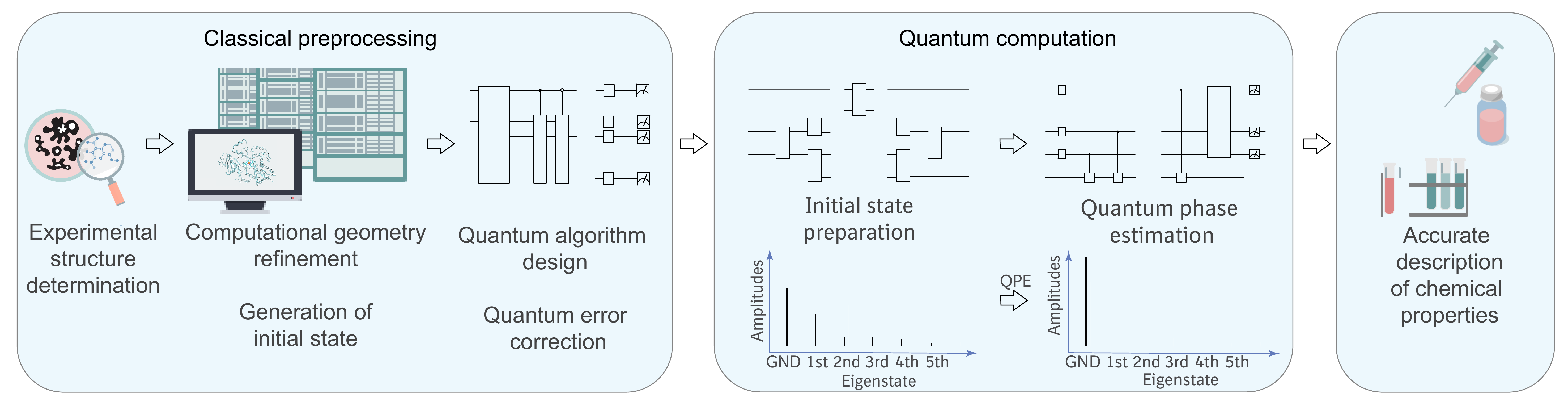}
    \caption{
    \protect\justifying 
    The workflow of electronic structure calculations on quantum computers using the quantum phase estimation (QPE) algorithm.  The first step includes classical preprocessing to optimise the geometry and the Hamiltonian. Afterwards, the quantum circuits are generated. The calculation on the quantum computer starts with the generation of the initial state, which is followed by the more expensive calculation of the ground state energy with the QPE. 
    The lower part of the quantum computation container:
    The initially prepared state consists of a superposition of many eigenstates but with a high overlap with the ground state.
    When the ground state (GND) energy is measured, the initial state is projected into the ground state.}
    \label{fig:calc_on_QC}
\end{figure*}

A quantum computer can perform such calculations in polynomial time without making any uncontrolled approximations if the initial state is close to the ground state~\cite{Berry2007, Ge2017, LinLin2020}. The ground state energy is computed with a combination of state preparation and quantum phase estimation (QPE). QPE is a very efficient algorithm to find the eigenstates and eigenvalues of a Hamiltonian, and it is at the core of many quantum computing methods. In Figure~\ref{fig:calc_on_QC}, we give an example of how these calculations can be performed on quantum computers for a chemical system. 
The presented workflow starts on a classical computer, which helps in refining the geometry of the chemical system, identifying a good initial state for the system and synthesising the error-corrected quantum circuit. The quantum computation starts with internally preparing this classically-determined initial state. The next step in the workflow is the application of QPE to the initial state. 
The cost of estimating the correct ground state energy depends directly on the overlap of the initial state with the ground state, and it becomes progressively more expensive as the overlap with the correct ground state decreases~\cite{kitaev_quantum_1995,Ge2017, LinLin2020, Garnet2022}. Modifications to this workflow allow for the calculation of other observables~\cite{Knill2007}, e.g., molecular forces~\cite{OBrien.npjQuantumInformation.2019, Sokolov2021, OBrien2021}.

Even though FTQC algorithms cannot yet be executed, many methods already exist to evaluate their computational cost. For example, for the ground state energy of the FeMoco~\cite{Reiher2017, Li2019}, through algorithmic improvements, the run-time estimates have been reduced from years to  days~\cite{Wecker2014, Reiher2017, Poulin2018, Lee2021, Blunt2022}. Further improvements will certainly come, and we will be able to perform such calculations in the future on an FTQC. In the next sections, we discuss the state of the art of drug design and where quantum computers could be employed to solve the electronic structure part of the problem for relevant pharmacological systems~\cite{Wecker2015, Tazhigulov2022, Blunt2022,Goings2022}.

\section{Computer-aided drug design}

\begin{figure}[tb]
    \centering
    \includegraphics[width=0.5\textwidth]{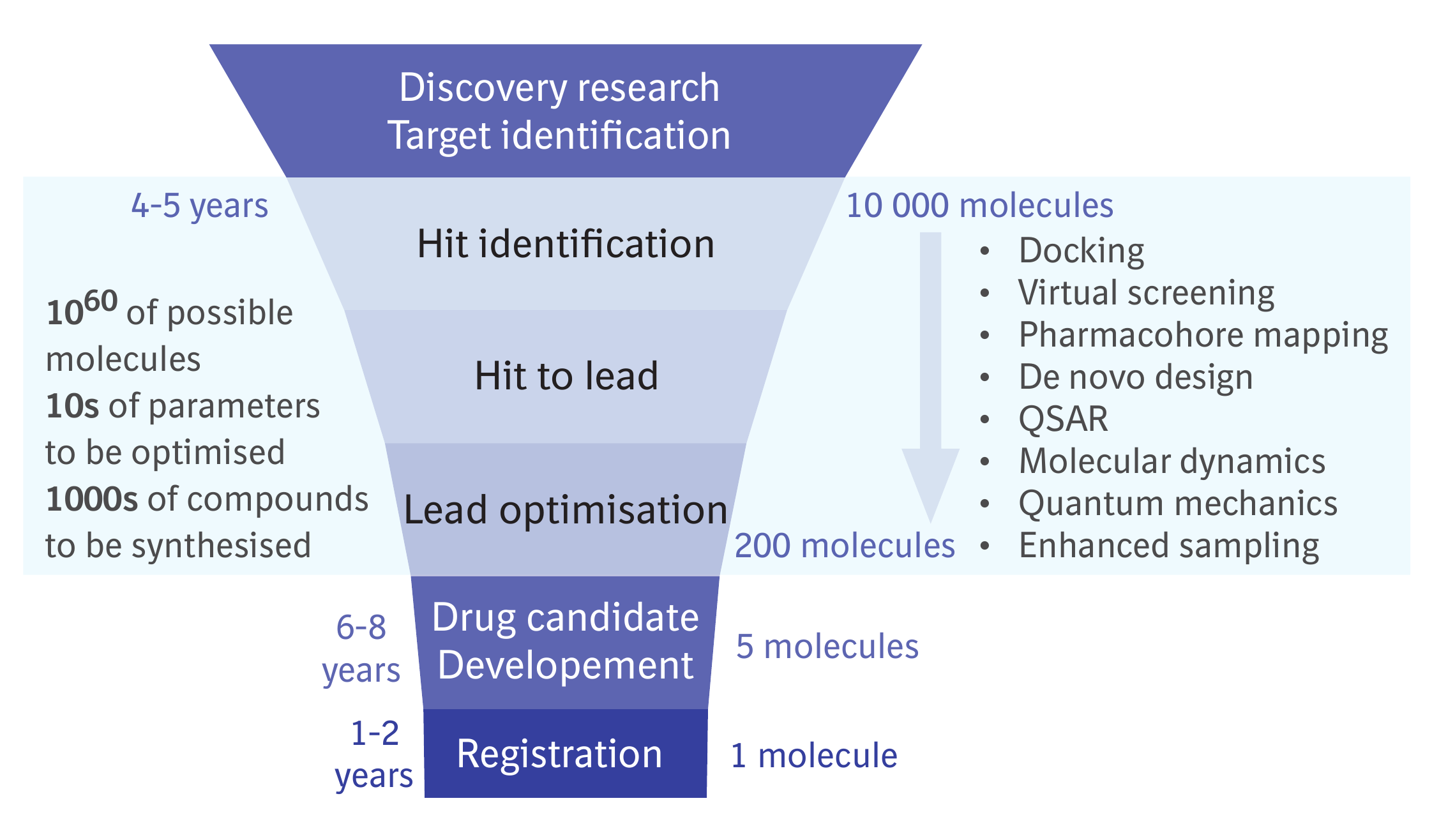}
    \caption{
    \protect\justifying 
    Workflow in the drug discovery process~\cite{Paul2010, Palermo2014}. Once the biological target has been identified, the process starts with the hit finding stage in a potential space of $10^{60}$ molecules~\cite{Polishchuk2013}. Through a repeated cycle of design, analysis, synthesis, and in-silico and in-vitro testing, the number of promising compounds is decreased from $10\,000$s to a few hundred by designing and selecting those with the best predicted and measured properties. Only very few highly optimised and safe molecules proceed into development towards the clinical trials, and only one is finally selected for approval by the medicinal agencies~\cite{Paul2010}. On the right side, the computational methods employed in the different stages of the drug design process from~\cite{Palermo2014} are listed.}
    \label{fig:Drug_design_cycle}
\end{figure}

Chemical compounds produced in the pharmaceutical industry result from a long process of discovery and refinement. The steps are summarised in Fig.~\ref{fig:Drug_design_cycle}. The drug discovery process starts with identifying a target protein involved in the disease pathology. Pharmacological modulation of this target is assumed beneficial for treating the disease~\cite{Hill2015, KENAKIN2017} and  is achieved with a molecule binding to the target. Identifying oral drug candidates, the most preferred form of drug administration, takes a long time, starting with very weak binders and taking several years of optimisation towards efficacious and safe molecules~\cite{Scannell2012, Palermo2014, Jayatunga2022}.

Millions of compounds are initially screened out of $10^{60}$ potential molecules~\cite{Polishchuk2013}. In the initial stages of the process, many different properties (e.g., binding affinity) have to be optimised. Therefore, in the so-called hit-to-lead and lead-optimisation programs, several thousands of molecules are synthesised before suitable candidates for the next steps towards clinical development are identified~\cite{Ferreira2017, Paul2010}.
Every synthesised molecule undergoes testing in-vitro (biochemical, biophysical, cellular), and in case of good properties, also in in-vivo (in an organism) assays; therefore, the goal is to achieve clinical candidates with the lowest number of optimisation cycles possible. In this phase of drug discovery, computational approaches are highly valuable by guiding the design of the right molecules, and recently several striking successes in computational design have been reported~\cite{Jayatunga2022, Allen2022}. 

Two major areas where computational chemistry can support drug design have been identified: 
(1) the prediction of pharmacokinetic properties (how the compound is absorbed, distributed, metabolised and excreted from the body), commonly realised by machine learning models trained on a wealth of experimental data from the heritage of projects in a pharma company~\cite{Talevi2018, Miljkovic2021,  maltarollo2015applying, Schneider2021}; 
(2) the calculation of the binding strength or binding affinity of a compound to the target, which is one of the most important properties of a drug candidate~\cite{Cournia2017, Deglmann2015}.  
The binding affinity is equivalent to the binding free energy between the drug and the target. It directly corresponds to the required local drug concentration at the target, determining drug efficacy. Therefore, it translates into the projected therapeutic human dose, the most important single parameter during drug design. Computations of the binding strength must be accurate in compound optimisation~\cite{Tinberg2013}. However, state-of-the-art methods based on molecular dynamics simulations with classical force fields do not perform reliably~\cite{King2021}. The goal is to achieve high accuracy (within 1.0~kcal/mol to experiment) because, at physiological temperatures, a 1.5~kcal/mol deviation already translates into a dose estimation which is wrong by one order of magnitude. On an atomistic scale, a system can be treated on a classical computer with many different levels of approximations for different sizes considered; see BOX~\ref{BOX:qcmethods} where some common methods are reported. 
In contrast to force fields, density functional theory (DFT) or coupled cluster (CC), which are methods based on quantum mechanics, lead to much better descriptions of molecular interactions but at a much higher computational cost~\cite{Deglmann2015}. 
\begin{table}[tb]
\refstepcounter{table}
\begin{tcolorbox}[colback=our-lightblue,colframe=our-blue,coltitle=black,title=Box \thetable: Common electronic structure methods employed on classical computers]
\begin{tabular}{p{35mm} p{35mm}}
\multicolumn{2}{p{76mm}}{
Commonly used quantum chemistry methods to solve the electronic structure problem.
In the left column, we zoom in on the Compound I intermediate of Cytochrome c Peroxidase (PDB ID: 1ZBZ~\cite{zbz,bonagura2003,berman2000}). As the method's accuracy increases from top to bottom, the molecules that can be calculated with classical hardware become increasingly smaller.
}\\ \\[-5mm]
\multirow{1}{*}{\begin{minipage}{34mm}\vspace{-10mm}
\centering
\includegraphics[width=25mm]{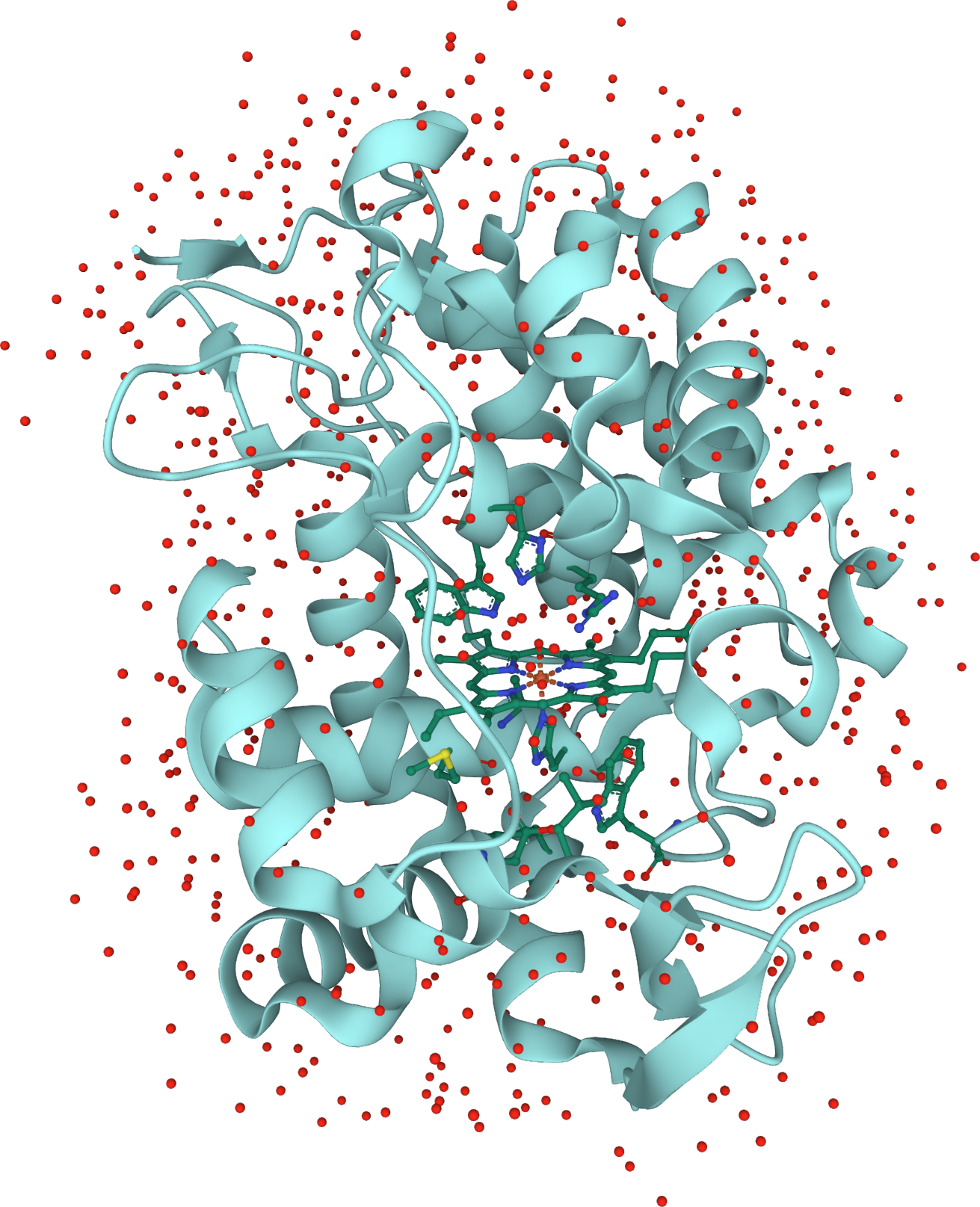}\\
Cytochrome c in solution
\includegraphics[width=25mm]{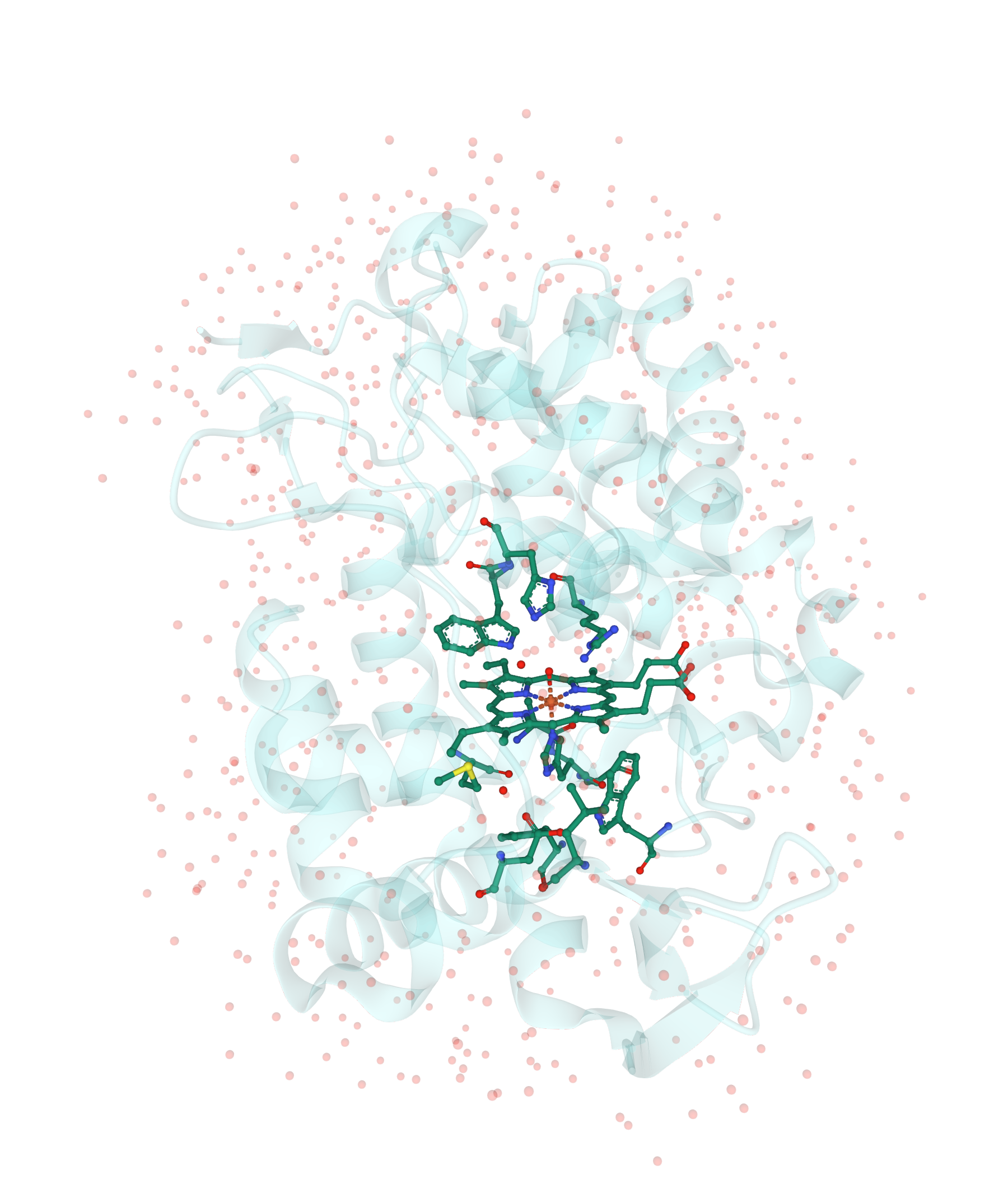}\\
Binding site
\includegraphics[width=25mm]{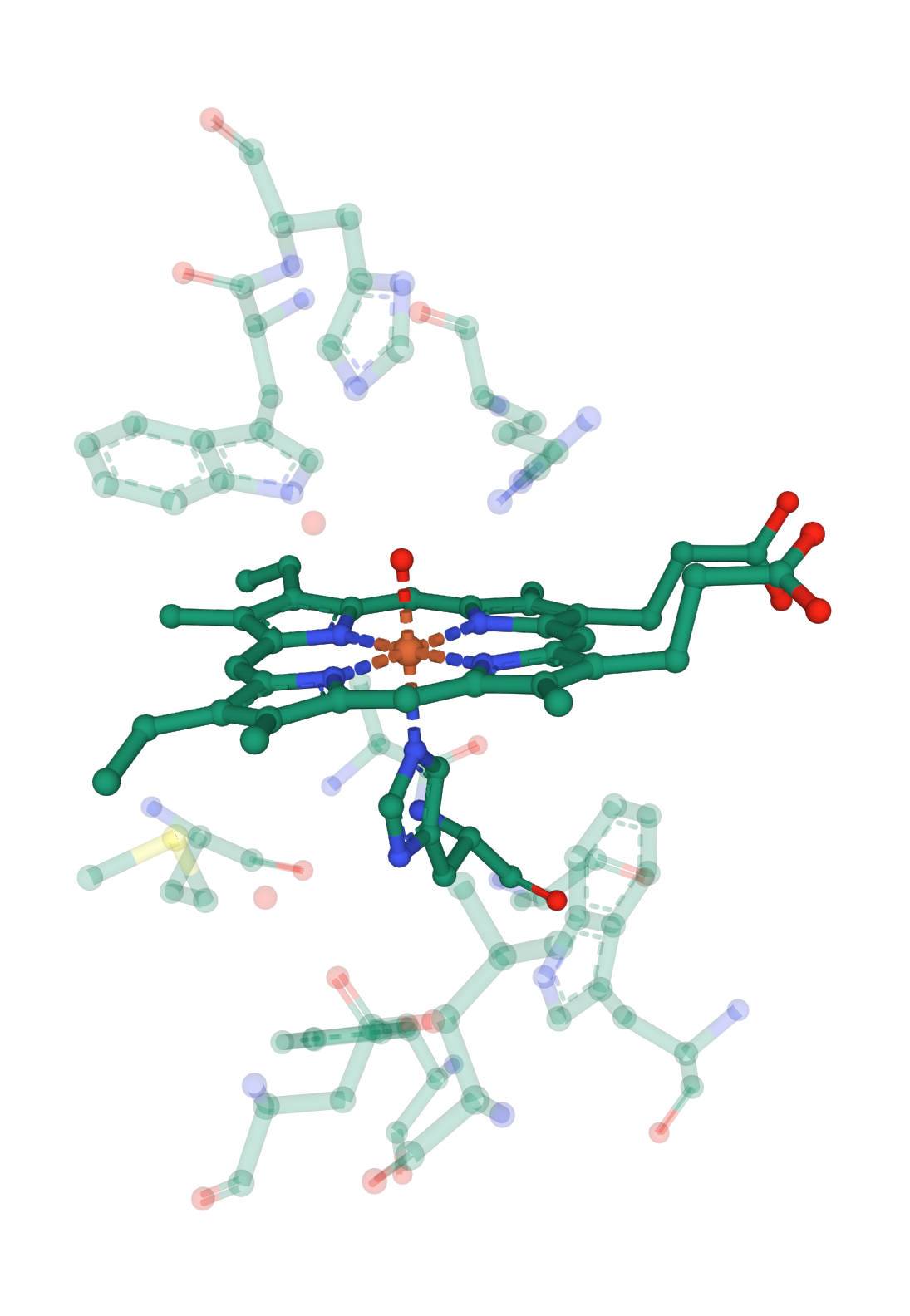}\\
Heme group
\includegraphics[width=25mm]{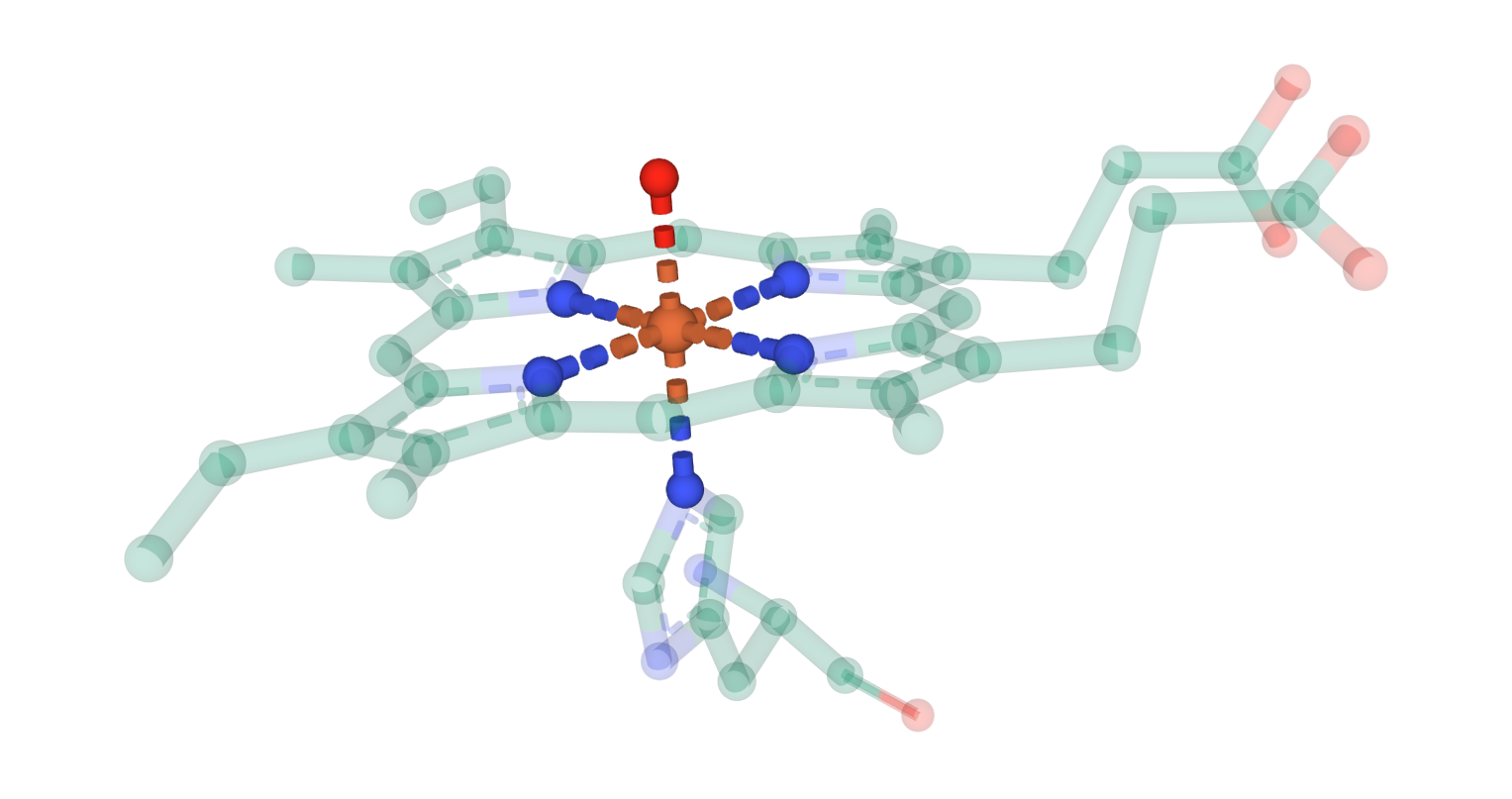}\\
Iron cluster
\end{minipage}} 
&
\begin{minipage}{40mm}\vspace{5.2mm}\flushleft\textbf{Force Fields/} \\
    \textbf{Semi-empirical Methods}\\
    \justifying
\noindent Methods that cannot fully describe quantum mechanical effects but can be tuned with information from quantum methods.
\end{minipage}
\\ \\[5.2mm]
&  
\begin{minipage}{40mm}\flushleft\textbf{Hartree-Fock/}\\
\textbf{Density Functional Theory (DFT)}\\
\justifying
\noindent
Mean-field methods treat electrons in the presence of the average potential of the other electrons. DFT includes electronic correlation, while Hartree-Fock does not. 
\end{minipage}
\\ \\[5.2mm]
&  
\begin{minipage}{40mm}\flushleft\textbf{Coupled-Cluster (CC)}\\
\justifying
\noindent
Cluster wavefunction methods that expand around a single mean-field reference.
 \end{minipage}
 \\ \\[6.6mm]
&  
\begin{minipage}{40mm}\flushleft\textbf{Full Configuration Interaction (FCI)}\\
\justifying
\noindent
Method that delivers the exact energy of the electronic structure problem within a finite basis set. \end{minipage}
\end{tabular}
\vspace{2mm}
\label{BOX:qcmethods}
\end{tcolorbox} 
\end{table}

\begin{figure}[tb]
    \centering
    \includegraphics[width=0.48\textwidth]{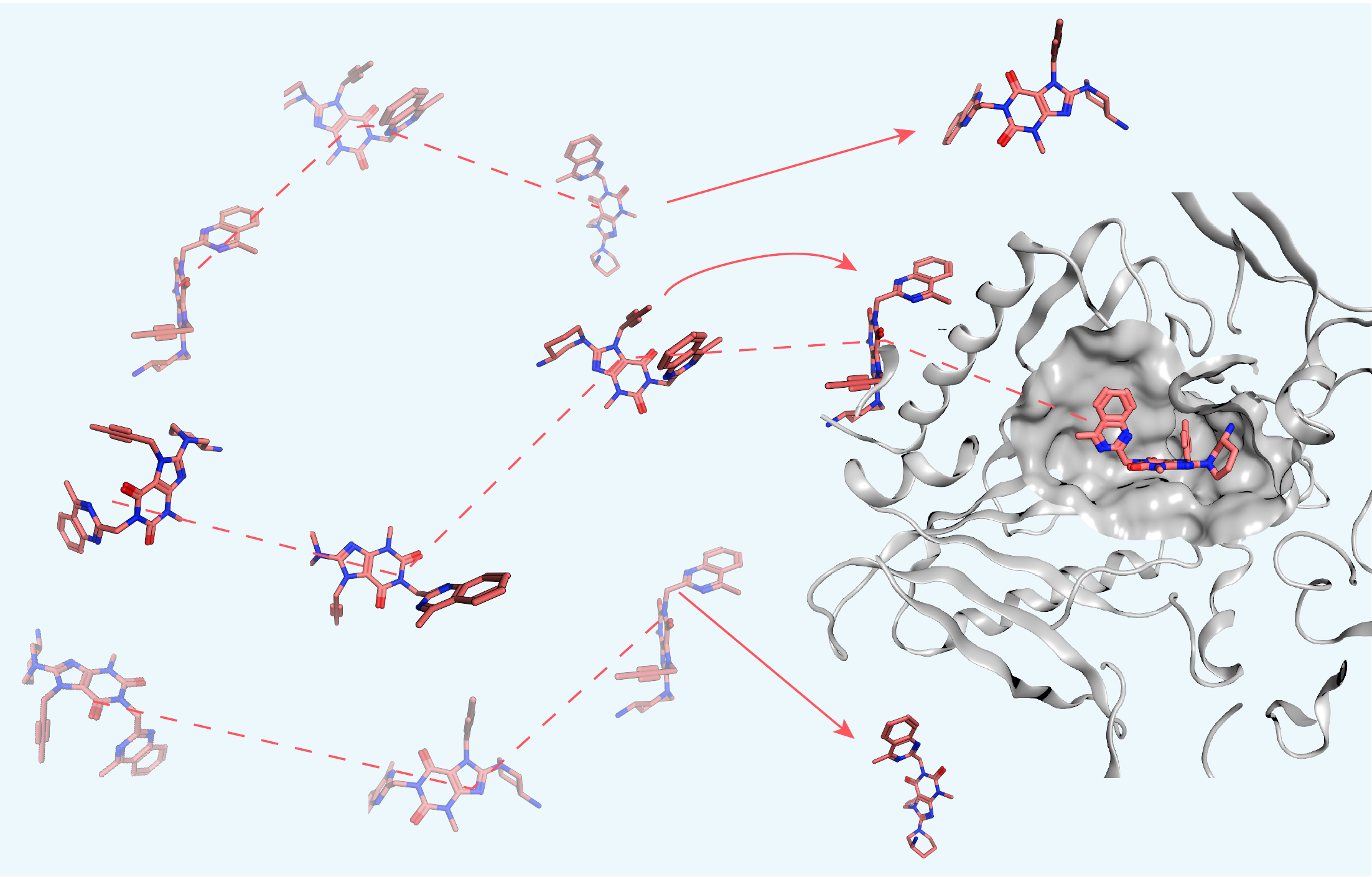}
    \caption{
    \protect\justifying 
    Schematic representation of a drug binding event (pdb ID: 2RGU). The ligand exists as an ensemble of conformations/geometries and orientations (left). Some approaches of the ligand towards the target result in binding, and some do not - as indicated by the arrows (right). Eventually, the sampling of ensembles of unbound and bound structures in solution yield the free energy of drug-target binding~\cite{Heilmann2020}. Equivalently, energy differences of ensembles of bound, structurally similar ligands directly relate to the difference in their binding strength.
    }
    \label{fig:binding_affinity}
\end{figure}

Other difficulties in these calculations stem from the thermodynamic nature of the compounds' properties~\cite{Heilmann2020, kaynak2022sampling}. A molecule can bind to a protein in many different ways~\cite{James2003}. One has to consider different accessible system geometries and binding pathways. One needs to identify the configuration with minimal free energy, which is the statistically most frequently observed one. In Figure~\ref{fig:binding_affinity}, the process of a molecule binding to a protein is pictorially depicted. Many different configurations and, thus, many single-point calculations must be computed.

When evaluating many compounds with similar chemical structures for their binding propensity, it is often faster to compute the difference in binding strength between the compounds directly; this task is often accomplished with {\em alchemical perturbation} methods~\cite{Cournia2017}, where a known compound is gradually morphed into a new one adapting the electronic structure accordingly~\cite{Song2020}. In this respect, simulating the ensemble properties, by, e.g.\ natural time-evolution, of the drug-target complex is a key step from which knowledge about thermodynamic properties can be directly derived~\cite{Karplus2002, vanGunsteren1998}.

The systems, including target, drug, and solvent, are made of several thousands of atoms (see BOX~\ref{BOX:qcmethods}), and free energy calculations require billions of single point calculations, where energy and force evaluation are performed, see Figure~\ref{fig:binding_affinity}. Furthermore, the necessary inclusion of explicit solvent (water) in the model can considerably increase the degrees of freedom and complexity~\cite{Hawkins2017,kaynak2022sampling}, making run-time often impractical. The calculation of the binding free energy of a small molecule to its target protein can take many hours on a classical computer. Increasing accuracy, e.g.\ by exploiting DFT, increases the calculation costs by several orders of magnitude, rendering the full DFT treatment for free energy calculations elusive. Higher levels of theory treatments, such as CC, which require even more computational resources, are, therefore, fully out of scope and can only be applied to small systems. 

Other potential use cases for quantum computing in drug development are the calculation and optimisation of reaction mechanisms~\cite{vonburg2021} for optimising the drug synthesis conditions and the calculation of molecular spectra for nuclear magnetic resonance (NMR), infra-red (IR) or vibrational circular dichroism (VCD) spectroscopy to identify structures~\cite{Joyce2017, gao2020general, Obrien2021NMR}.
However, the impact of quantum computing on these use cases for drug design would be rather modest if compared to the potential impact of better and faster calculations at the drug design stage (lead optimisation). 
For example, usually, drug synthesis costs are not the main driver of non-generic drug market prices. The reason for this is the economic need to balance out a large amount of failed optimisation programs and clinical trials, see~\cite{Paul2010}. Additionally, for the prediction of NMR spectra, lower accuracy methods such as DFT have been shown to achieve good results in many cases~\cite{buhl1999dft, Xin2017, gao2020general}

In summary, most of the use cases of quantum mechanical calculations in drug design would benefit from speed-ups to DFT and CC methods, which are still too slow for broader application in the drug development process but offer good-enough accuracy for most systems. This is because most oral drugs are  small closed-shell organic molecules (they need to pass through the gut wall to be absorbed) which generally lack strong correlation and, with some rare exceptions, e.g.\ cytochrome P450 interactions in drug metabolism~\cite{Goings2022},  can be treated with lower accuracy methods due to their general elemental compositions~\cite{Smith2014}. However, few examples of drug molecules with metal centres exist, for example, for cancer treatments or contrast-enhanced imaging of tissues~\cite{Phillips2019}. An open, unexplored question is whether this scarcity of potentially strong-correlated drugs is due to some intrinsic unwanted features of metal-bearing drugs. This could lie in their undesired pharmacokinetic behaviour or potential toxicity, which would make them unfit as drugs. A different possibility is that they have been avoided due to the challenges in their computational optimisation.

\section{Challenges and prospects}

The current limitations of quantum chemistry in drug design either come from a lack of accuracy (for the few described difficult systems) or the large computational costs of the DFT calculations for ensembles of bio-molecules. For both limitations, quantum computers do not give an immediate remedy yet, although promising ideas are starting to emerge.

Currently, quantum computers are expected to speed up electronic structure calculations for strongly correlated systems with already-known quantum algorithms (e.g.\ QPE). This could be used, for example, to better understand the physics of cytochrome P450~\cite{Goings2022}. However, the largest impact will come if one can go beyond calculating single-point energies of strongly correlated systems. 

The last 30 years have seen dramatic improvements, both on the hardware side, as well as on the algorithmic one~\cite{Lee_2017, Campbell2017, Kim2022, Fowler2012,vonburg2021, Berry2019, Reiher2017, Liu2022,  arute2019quantum, Wu2021, madsen2022quantum}. 
Even though these improvements have enabled the impressive quantum computing capabilities we have today, there is much more needed to make quantum computing practical for drug discovery.

Concerning the run-time of algorithms, quantum error correction represents one of the dominant sources of overhead costs in space and time for executing fault-tolerant quantum algorithms. Error correction requires thousands of physical qubits for each logical qubit~\cite{Fowler2012}, resulting in millions of qubits for calculating the FeMoco ground state energy~\cite{Reiher2017, Lee2021}. To reduce these overheads, not only better hardware with lower error rates and increased qubit connectivity needs to be developed but also new further improvements to quantum error correction should be explored~\cite{Lee_2017, Campbell2017, Kim2022}. 
 
On the algorithmic side, one of the central yet unresolved challenges is preparing an initial state because the run-time of QPE directly depends on this state. Even though the run-time has improved over time~\cite{Ge2017, LinLin2020}, the dependence on the overlap of the initial and target states cannot be circumvented~\cite{kitaev_quantum_1995}. Several heuristic solutions have been proposed~\cite{Aspuru2005, Wecker2015, Tubman2018}, while further studies are required to understand the extent of this problem fully. For the case of weakly correlated systems, a potential solution relies on decomposing the system into smaller sub-systems and applying a series of QPEs on these to maintain the overall overlap.

Another essential research direction is the reduction of the overall computational cost by, for example, finding more compact representations of the systems' Hamiltonian, which directly impacts the run-time of the quantum algorithms~\cite{Barcza2011, Lee2021, von2021quantum}.
At the same time, analogously to classical algorithms, it should be possible to find quantum algorithms for specific cases based on heuristics that scale much better than general algorithms. Yet the absence of error-corrected quantum computers prevents the thorough benchmarking of heuristics today. However, there might be more systematic approaches to analyse the scaling and constant factors of heuristics for specific input parameters.

Current quantum algorithms focus on delivering speed-ups at the highest accuracy, which is not always relevant for industrial applications. Substantial run-time improvements compared to approximate classical methods would have a more considerable mid-term impact. However, speeding up approximate techniques on a quantum computer seems quite challenging. DFT and Hartree-Fock already have linear scaling implementations on classical computers, and it will be difficult to outperform them on a quantum computer. Instead, a quantum computer could provide new insights into the systems' physics to improve the classical methods. For example, we could use quantum computations to design better functionals for DFT. Alternatively, it might be viable to use quantum computers to speed up classical calculations in contracting tensor networks~\cite{Haghshenas2022, Kim2017}. Implementing CC methods on quantum computers could achieve a quadratic speed-up for the optimisation phase~\cite{Gilyen2017}. 
Another possibility is to save computational cost by exploiting perturbation theory on quantum computers~\cite{Mitarai2022}. Recent results have also shown that quantum computers can outperform classical mean field methods in simulating electron dynamics~\cite{Babbush2023}.
In the future, one could explore new routes in finding a trade-off between accuracy and costs, for example, by tuning the numerical accuracy of the Hamiltonian simulation~\cite{Abrams1999} or by truncating the amount of information in the Hamiltonian.

On the drug design side, while single-point calculations can give insights into systems' physics, we typically require billions of single-point calculations to determine thermodynamic quantities, e.g.\ binding affinity. This large number of calculations, combined with the quantum computing run-time on the order of days for one of them~\cite{Goings2022, OBrien2021}, makes it impossible to obtain results in a reasonable time, let alone compete with run-times of highly optimised experiments. 
A potential route to a more practical calculation of thermodynamic quantities might come from simultaneously modelling classical nuclei  and electrons in one wave function on the quantum computer. One can envisage calculating thermodynamic properties, e.g., the free energy, directly on a quantum computer by generating thermal ensembles of geometries~\cite{Somma2007}. Additionally, treating the nuclei quantum mechanically would help interpret molecular spectra~\cite{dinu2019toward}.

On a more speculative side, quantum machine learning algorithms applied to the outcome of quantum computations have the potential to predict pharmacokinetic properties~\cite{Aleksic2022, Carleo2019}. When large quantum computers become available, we might be able to compute wave functions of many ensembles of molecules and subsequently run quantum machine learning algorithms on these wave functions~\cite{Biamonte2017, Huang2021, McClean2021, Huang2020}. 

\section*{Conclusion}

Current classical computing methods fail to describe quantum systems accurately enough in relevant times for the pharmaceutical industry, limiting the applicability of quantum chemistry to drug design. More accurate computations could bring significant value to the pharmaceutical industry by replacing many labour-intensive experiments with calculations \textit{in silico}, as long as the computational cost is lower than the experimental effort. Quantum computations could enable key, experimentally inaccessible insights into chemical systems, exploiting methods that directly derive properties from wave functions~\cite{McClean2021}.
 
To have a profound impact on the pharmaceutical industry, quantum computers need to benefit a broader set of problems than the small number inaccessible to classical computers~\cite{Liu2022, Cao2019}. Typical relevant systems have thousands of atoms, e.g.\ large protein structures with their surroundings, and rarely require exact accuracy. However, in many pharmaceutical use cases, one must determine thermodynamic properties that rely on large thermodynamic ensembles, thus requiring many single-point calculations.
Finding new methods that allow trade-off accuracy for time on quantum computers or that avoid sampling could be beneficial.
Ideally, quantum computers should offer accuracy and robustness for both strongly and weakly correlated systems at a speed that is currently only accessible by lower-accuracy methods. By getting rid of some of the current approximations, quantum calculations in drug design would become truly predictive and much more widely used. 

Major advancements in quantum algorithms for electronic structure problems brought down computational costs~\cite{Hastings2015, Low2019, Lee2021, vonburg2021} over the last years, while further improvements are required for practical applications in industry. Furthermore, fundamental improvements in hardware, error correction codes and algorithms (e.g.\ for state preparation) are necessary to go beyond single-point energy calculations. 

Steps are already being made towards solving some of these challenges, and several routes exist to achieve these goals. 
We are convinced that open research integrating academia and industry will help make quantum computing an essential tool to design better drugs faster.

\section*{Acknowledgments}

The authors thank Darryl McConnel, Alexander Renner, Christoph Ehrendorfer, Lorenzo Pautasso, Manuel Möller, and Anika Pflanzer for comments on the various iterations this perspective went through. The molecules reported are visualised in {\tt Mol*}~\cite{sehnal2021}.

\bibliography{main}

\end{document}